\documentclass[letter]{jpsj2}
\usepackage{latexsym}
\usepackage{amsmath}
\usepackage{bm}
\usepackage{color}

\newcommand{\udarrow}[2]{\smash{\mathop{%
  \hbox to 0.6cm{$\rightleftharpoons$}}\limits^{#1}\limits_{#2}}}

\title{%
Chiral Crystal Growth under Grinding}

\author{%
Yukio \textsc{Saito}\thanks{yukio@rk.phys.keio.ac.jp}
 and Hiroyuki \textsc{Hyuga}\thanks{hyuga@rk.phys.keio.ac.jp}
}

\inst{%
Department of Physics, Keio University, Yokohama 223-8522 
}

\recdate{\today}

\abst{%
To study the establishment of homochirality observed in the 
crystal growth experiment 
of chiral molecules from a solution under grinding,
we extend the lattice gas model of crystal growth as follows.
A lattice site can be occupied by a chiral molecule in R or S form, 
or can be empty.
Molecules form homoclusters by nearest neighbor bonds.
They change their chirality if they are isolated monomers
in the solution.
Grinding is incorporated by cutting and shafling the system randomly.
It is shown that Ostwald ripening without grinding is extremely slow
to select chirality, if possible. 
Grinding alone also cannot achieve chirality selection.
For the accomplishment of homochirality, we need 
an enhanced chirality change on crystalline surface. With
this "autocatalytic effect" and the recycling of monomers 
due to grinding, an exponential increase of
crystal enantiomeric excess to homochiral state is realized.
}

\kword{%
homochirality, crystal growth, lattice gas model, grinding, 
kinetic Monte carlo simulation, autocatalysis, recycling
}

\begin{document}
\sloppy
\maketitle


It has long been known that organic molecules in life
is one-handed: Among the two possible enantiomers with 
mirror-symmetric stereostructures,
amino acids choose L-form and sugars D-form \cite{pasteur48,japp98}.
To understand the possible origin of this handedness or homochirality, 
Frank proposed a model rate equation with nonlinear reaction terms
\cite{frank53}.
In a field of organic chemistry, Soai and his group found, for the first
time, an organic system which shows the amplification of enantiomeric excess (ee)
in the production of chiral molecules from achiral substrates
\cite{soai+95}, but the system cannot reach a unique state 
with homochirality:
The final state and its ee value depend on initial conditions.
The ee amplification is explained by the second order nonlinearity
in the rate equation. It is induced by an autocatalytic effect of
 homodimers in the preference of their enantiomer production \cite{sato+03}. 
We have recently shown that if a recycling path is incorporated
such that the products decompose into achiral substrate, 
a homochiral state is possible to realize as a final state
 \cite{saito+04}.
Recently, the ee amplification is also observed in another
chemical system more relevant to life \cite{mauksch+07}.

Chirality of crystals is also known  for a long time in their shapes
and in their optical activity:
When achial silicon dioxide (SiO$_2$) molecules build up in a crystal of quarz,
 it has two forms which are the mirror images of each other.
It is a well known fact that Pastuer separated salt of tartaic acid
by the differcence in crystalline shape, and discovered molecular optical
activity \cite{pasteur48}.
An achiral molecule, sodium chlorate (NaClO$_3$), also crystallizes in two kinds of
enantiomeric
form from its solution, and has been studied for a long time,
as a prototype of chiral symmetry breaking
\cite{kipping+98,kondepudi+90,viedma04,noorduin+08a,noorduin+08b}.
In a Viedma's experiment, a mixed solution of 
two types of crystallites was continuously ground by glass balls, and it then
turned out that a unique chirality was selected; that is, the
homochirality is achieved \cite{viedma04}. 
In a very recent experiment by Noorduin {\it et al.}, crystals 
of chiral molecules are grown under constant grinding,
and the conversion of molecular chirality is found \cite{noorduin+08a}.

It is easily recognized that the grinding corresponds to the recycling process
in the reaction model. It is, however, not obvious what kind of process
in the crystal growth plays the role of the nonlinear production.
In a standard picture of cluster dynamics for the crystal growth
\cite{becker+35,saito96},
there is a size dependence of the growth and dissolution rates, which
leads to Ostwald ripening \cite{ostwald97}, but nonlinear
effect is not included,
Therefore, in the theoretical attempts to explain the homochirality
in crystal growth, one imposes rather arbitrarily
a parallel incoporation of both  small chiral units
 and achiral substrates into large crystal clusters to
provoke nonlinear effects mathematically
\cite{uwaha04,uwaha08,saito+05}.

Here we present a novel microscopic picture for the physical processes 
taking place during the growth of chiral crystals,
which leads to the nonlinear process.
Corresponding to Noorduin {\it et al.}'s experiment\cite{noorduin+08a},
 only chiral molecules are considered. Molecules of the same enantiomeric type
makes chemical bonds and form homoclusters, which grow into crystallites.
Enantiomers also mutually convert their chirality, if they 
are in an isolated 
monomer state in solution without forming homoclusters. 
We further assume that,  as the number of neighboring enantiomers 
of opposite type increases, the corresponding chirality 
conversion rate is enhanced from that of a completely isolated monomer. 
For instance, the chirality of one enantiomer is more easily changed on the 
surface of the crystal of another enantiomer
than in a middle of the solution. Then, after the chirality change
they are quickly incorporated
into the crystal. It looks similar to the autocatalytic
process in chemical reaction.
Once we accept this "autocatalytic" process, 
grinding not only provides us a recycling of monomers back from
clusters, but also increases numbers of surface and 
kink sites to enhance nonlinear autocatalytic effect. 
In order to demonstrate the feasibility of this senario, we construct a 
toy model of a lattice gas system, and study on the chirality selection
by means of kinetic Monte Carlo (KMC) simulations.\\

\noindent
{\it Lattice Gas Model for Crystal Growth under Grinding}

We construct a lattice gas model of crystals  
composed of chiral molecules, growing from a solution, 
but also constantly
broken into small fragments by grinding.
The system consists of a square lattice with a size $L^2$ where
a lattice constant being set to unity.
Periodic boundary condition is imposed during the crystal growth
both in $x$ and $y$ directions, where $x$ and $y$ directions are parallel to 
the perpendicular edges of the square lattice, respectively.
On a lattice site there can be a molecule or a site can be empty.
Molecule takes two enantiomeric form, which we call R and S,
 for short. Only the same type of enantiomers is assumed to form
crystal clusters. These homo-clusters grow
 by incorporating monomer molecules. 
When a R molecule makes contact with another R monomer or
with R-crystal clusters, it is instantaneously
incorporated into a homocluster.
The same holds for a S molecules to S clusters.

An isolated monomer changes its site with a rate $k$.
If the new site is next to the original one, the 
motion leads to the diffusion of monomers.
In the cases of relevant experiments 
\cite{viedma04,noorduin+08a,noorduin+08b}, 
however, fully grown crystals have 
facets, and the growth is seemingly limited by surface
kinetics, but not by diffusion\cite{saito96}.
Furthermore, we want to study the crystal growth under grinding 
where the solution is strongly stirred.
Forced convection wipes out the effect 
of chemical diffusion. Therefore, we assume molecules
 change their sites to far distant ones:
A new site for an isolated monomer is 
anywhere on an empty lattice site.
If the jumped site is next to the cluster of the same enantiomer,
the  monomer is incorporated into the cluster.

Only with the above incorporation process, the grown crystal
clusters have a ramified shape. For those surface molecules 
poorly coordinated to the crystal, thermal excitation 
leads their detachment.
For a monomer to detach from a homocluster, it has to break chemical
bonds, whose energy is set $J$ per bond, 
Then, the rate of detachment from a site with $n$ nearest neighbors
of the same type of enantiomer
can be assumed
\begin{align}
k e^{-nJ/k_BT} ,
\label{eq1}
\end{align}
at a temperature $T$ with the Boltzman constant $k_B$.
After the detachment, the monomer has to be placed again anywhere
on an empty site to conserve the total number of molecules.
By assuming that a temperature is low enough, we neglect
detachment process from those sites coordinated higher than the
kink site, namely, detachment rates  for $n= 3$ and 4 vanish.

It is known that monomers
convert their chirality in a solution \cite{noorduin+08a}. 
We denote the conversion rate as $\nu_0$;
\begin{align}
\nu_0: R \rightleftharpoons S ,
\label{eq1}
\end{align}
when monomers are isolated.
In order to incorporate the autocatalytic effect in the crystal
growth process, we assume that when a chiral monomer is
surrounded by its opposite enantiomers, conversion rate is enhanced
by some factor. 
For example, when a R monomer is at the step edge 
of an S-crystallite with a single S neighbor, 
the conversion rate is $\nu_1$, which is greater than $\nu_0$.
When it is at the kink site of an S-crystallite with two S neighbors,
the rate is $\nu_2$, and so on.
Similar autocatalytic processes are used in the study of
spatial propagation of homochirality in the case of chemical reaction system
\cite{saito+04b,brandenburg+04,shibata+06}.

In each KMC step, time increases by 
the inverse of the total transition probability, 
and after every time interval $f^{-1}$ the following grinding 
process is introduced.
Grinding process leads not only to a long-ranged positional 
change of all the monomers and clusters, but also
to the breaking of growing crystallites which
provides multiples of steps and kinks.
In order to mimic this ablation effect, we set aside 
periodic boundary conditions, and
cut the system randomly
into two parts by a line parallel to $x$ (or $y$ ) axis.
Then, we reflect one part
through its central axis, and exchange the position of two parts. 
Only with this vertical or horizontal cut and shafling,
no appreciable enhancement of kink sites are expected.
Therefore, we introduce in addition diagonal cuts;
cut the system by two diagonals and exchange left triangle
with the right one, or upper triangle with the lower one,
randomly.
If the diagonal cut runs through the crystal cluster,
a large amount of new kink sites will be created.
Because of the random shafling in vertical or horizontal directions,
the center of the diagonal cut shifts effectively in random.
The combination of random cuts is repeatedly applied in the KMC simulation 
with a frequency $f$:
The frequency $f$ is a measure for the strength of grinding.\\

\noindent
{\it Ostwald Ripening}

If the grinding does not take place, crystal clusters grow in size.
In case with only one species of molecules, crystals of different
sizes compete with each other to incorporate monomers, and eventually
the largest one wins.
This coarsening phenomena is called Ostwald ripening
\cite{ostwald97}.
As we have two types of enantiomers, two largest crystallites made 
of R and S molecules remain eventually, and they compete through the process
of chirality change in solution.
This is, however, a very slow process \cite{noorduin+08b}, 
depending on the chirality conversion rate
$\nu_0$. These expected features are confirmed in numerical simulations as 
is explained below.

In all the numerical simulations to be presented, 
the system size is set $L^2=100^2$
and a total concentration of molecules is $c=0.1$, so that 1000 molecules
participate in crystal growth.
The rate of monomer positional change is set $k=10^4 \nu_0$, whereas 
the chirality conversion rate $\nu_0$ is much smaller than $k$.
In this section, we study the case without any autocatalytic effect so that 
we set $\nu_1=\nu_2=\nu_0$. 
Since $\nu_0 \ll k$,
an isolated monomer does not change its chirality so often
compared to the positional change.
The Boltzmann factor for the detachment
is set $\exp(-J/k_BT)=10^{-3}$, which is small enough to achieve 
polygonal crystals with facets, but also large enough to 
observe the Ostwald ripening phenomena in a relatively short
simulation time.

\begin{figure}[tbh]
\begin{center} 
\includegraphics[width=0.3\linewidth]{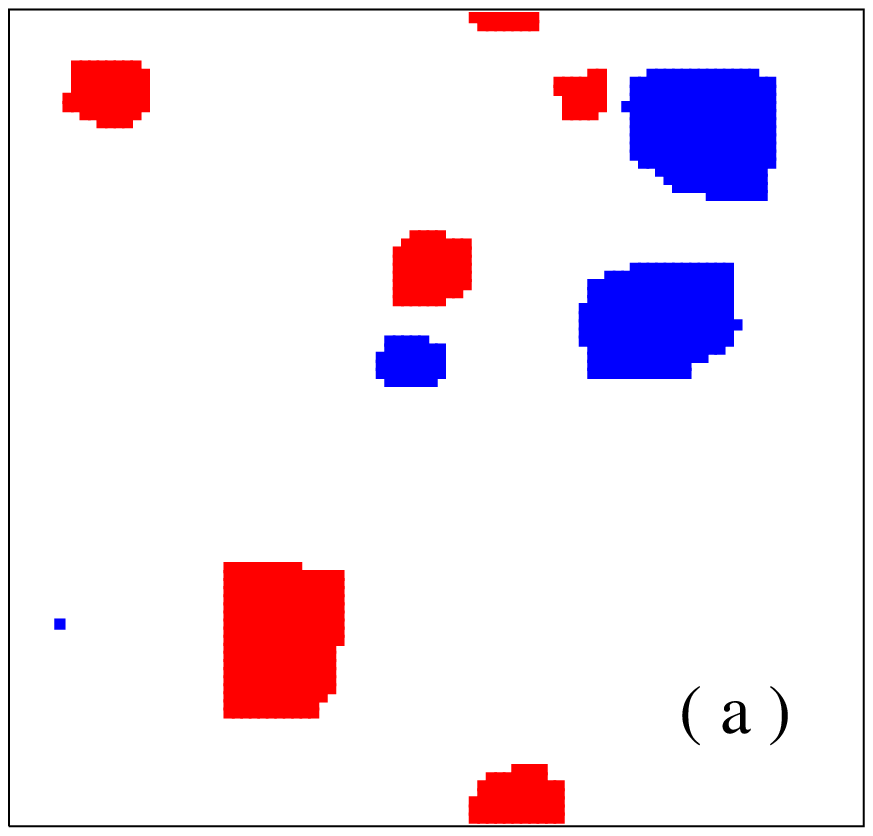}
\includegraphics[width=0.3\linewidth]{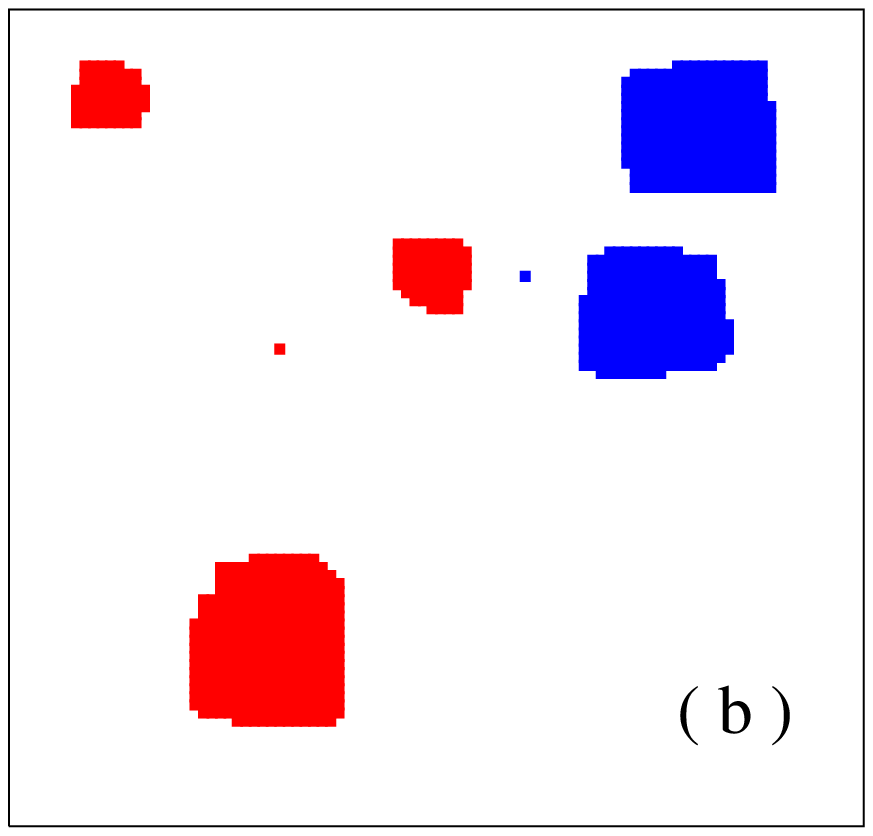}
\includegraphics[width=0.3\linewidth]{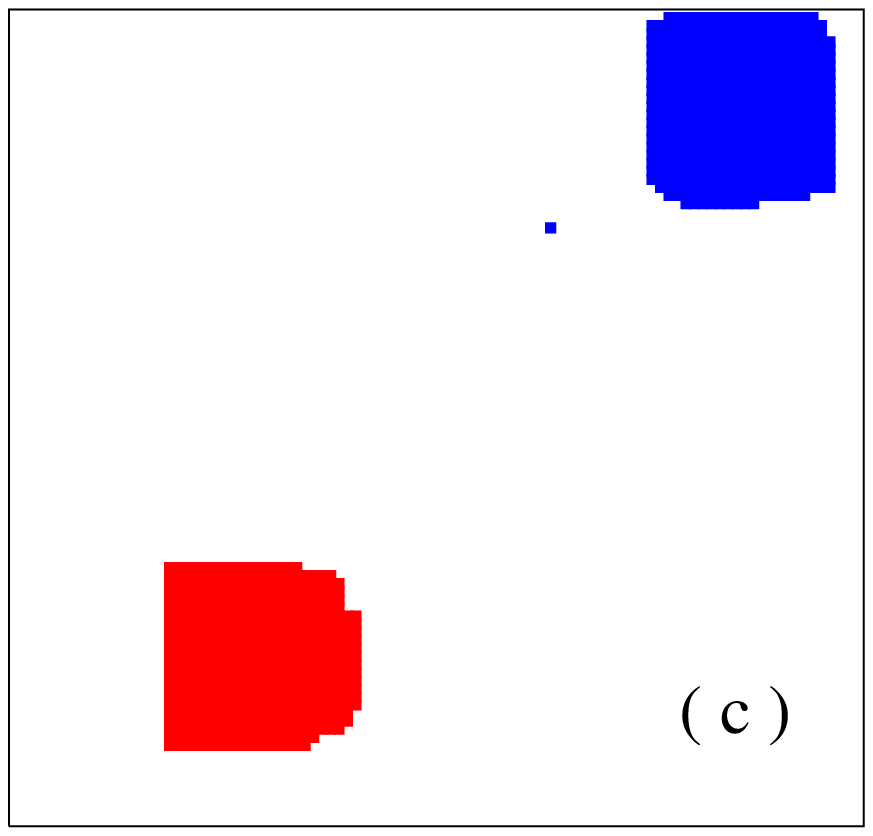}
\end{center} 
\caption{Ostwald ripening: Configurations of crystal clusters
at times (a) $\nu_0 t=4240$, (b) 10850, and (c) 73,000.
Red and blue color correspond R and S molecules, respectively.
}
\label{fig1}
\end{figure}

In Fig.1, we show a simulation result for the case where there are the same
number of R and S molecules; $N_R=N_S=500$.
Since we have a small closed system with few numbers of molecules,
the number of clusters decreases to two, as is shown in Fig.1
after a long time, $\nu_0 t=73,000$.
If one waits much longer until a single cluster wins in the competition 
of Ostwald ripening, 100\% of crystalline enantiomeric  excess
 will be established.
However, in an actual experiment, no coarsening is observed in a
realistic time scale \cite{noorduin+08b}.
There is no sufficient time to reach a 
final state with a single or a very few crystal clusters
in an actual experiment.

The degree of chiral symmetry breaking is measured by a
chiral order parameter
\begin{align}
\phi=\frac{N_R-N_S}{N_R+N_S}
\end{align}
for a system with the numbers of R and S molecules $N_R$ and $N_S$,
respectively, 
or by its absolute value $|\phi|$, normally called crystaline 
enantiomeric excess (cee).
It varies very slowly as shown by a red curve, A1, in Fig. 2(a), and
even its sign can change from time to time.
$|\phi|$ is fluctuating but remains small through the
whole simulation time till $\nu_0 t=73,000$( not shown in Fig.2).

Even when the initial state has a finite chirality, as cee=0.2,
the evolution of the chirality is very slow, as shown by the curve 
A2 in Fig. 2(a). In some case, after the coarsening of each enantiomer is
over with only two competing clusters remaining, a slow conversion to the
dominant enantiomer takes place, as expected from the Ostwald ripening.
The process proceeds very slowly, and even at $\nu_0 t=7 \times 10^4$, 
final homochiral state is not achieved.\\

\begin{figure}[tbh]
\begin{center} 
\includegraphics[width=0.45\linewidth]{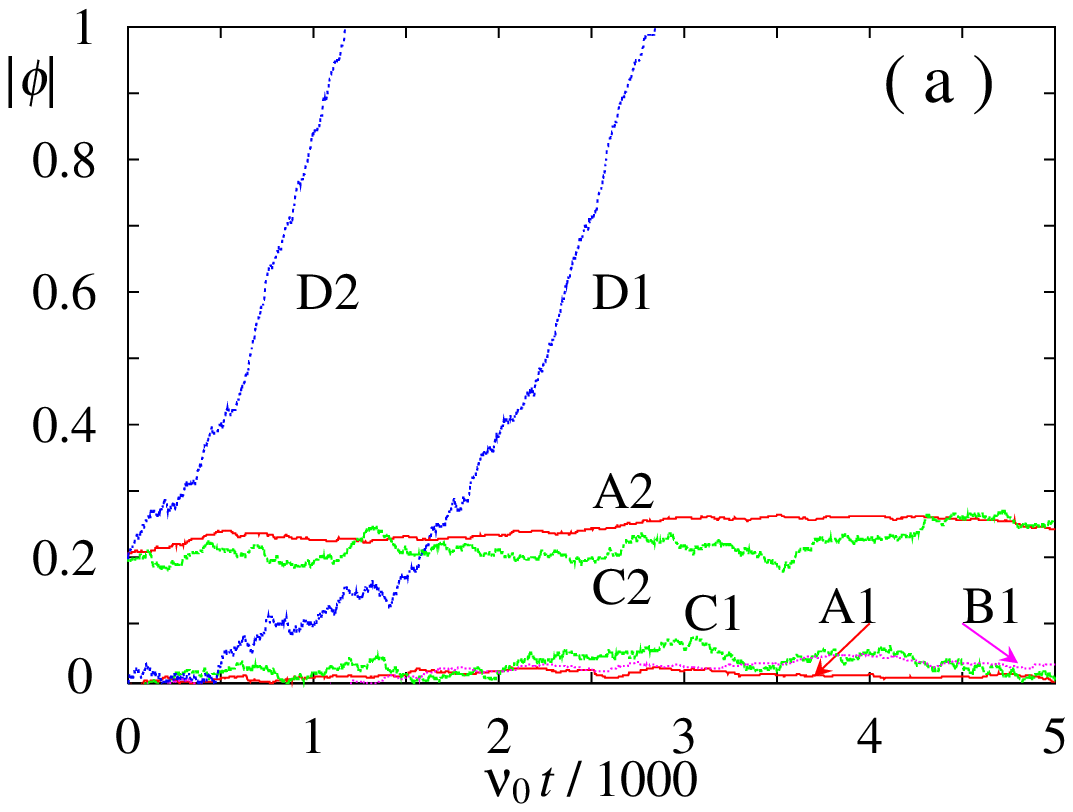}
\includegraphics[width=0.45\linewidth]{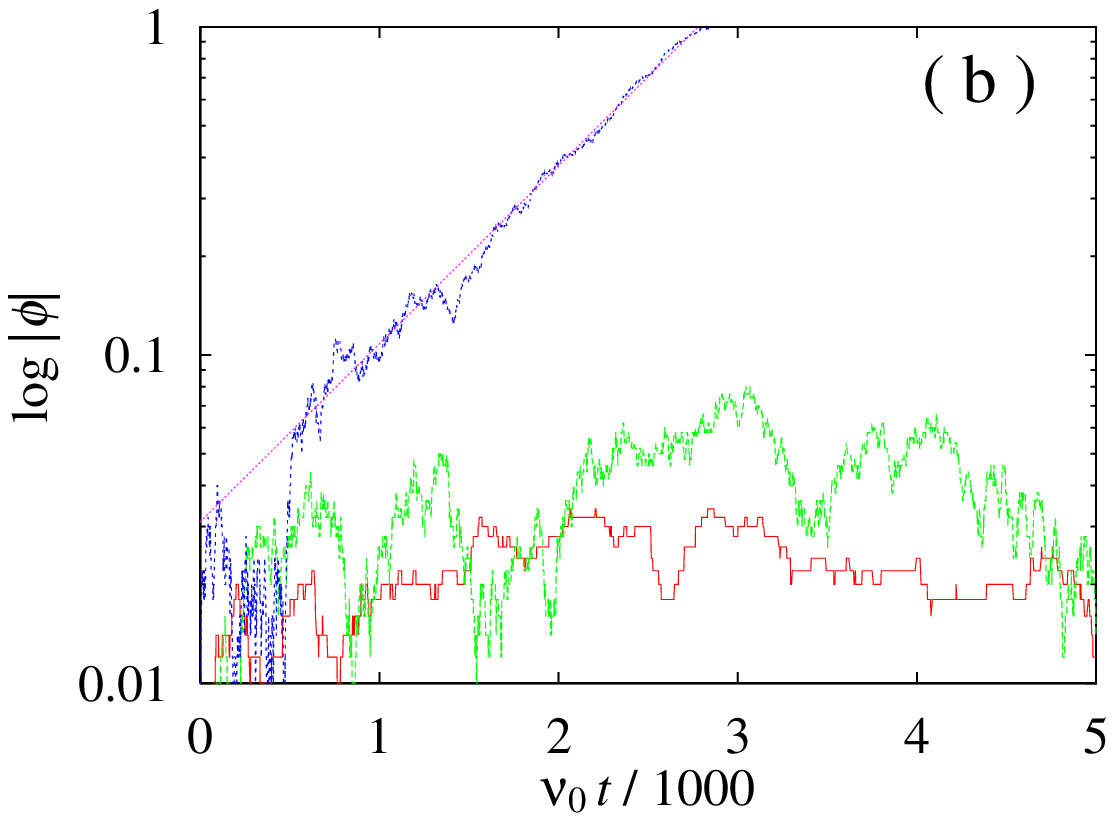}
\end{center} 
\caption{Absolute value of the order parameter $|\phi|$, that is cee,
 versus elapsed time for
different cases; (a) curves A (red) correspond to the case with
neither grinding nor autocatalysis and thus to Ostwald ripening,
curves B (pink) to the case without grinding but with autocatalysis,
curves C (green) to the case with grinding but without autocatalysis, 
curves D (blue) to the case with both grinding and autocatalysis.
Gringing frequency is $f/\nu_0=1$.
Curves with index 1 represent results from a racemic
 initial condition, and ones with index 2 represent result for the case
 that the initial state is partially chiral with cee=0.2.
(b) $\log |\phi|$ versus elapsed time started 
from a racemic initial state. 
}
\label{fig2}
\end{figure}

\begin{figure}[tbh]
\begin{center} 
\includegraphics[width=0.3\linewidth]{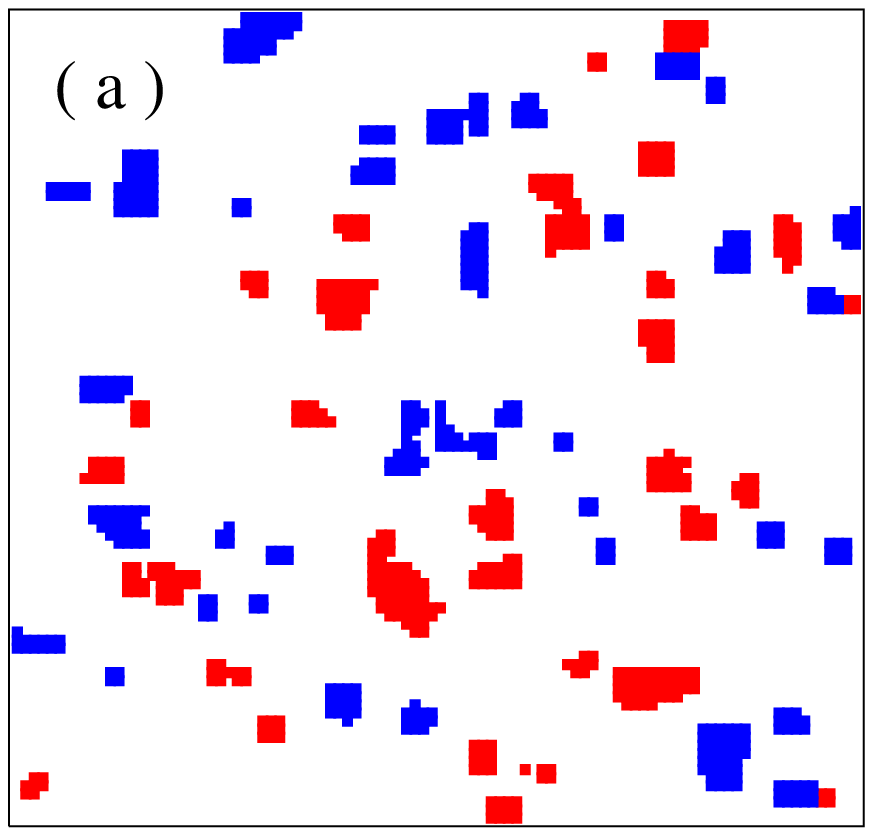}
\includegraphics[width=0.3\linewidth]{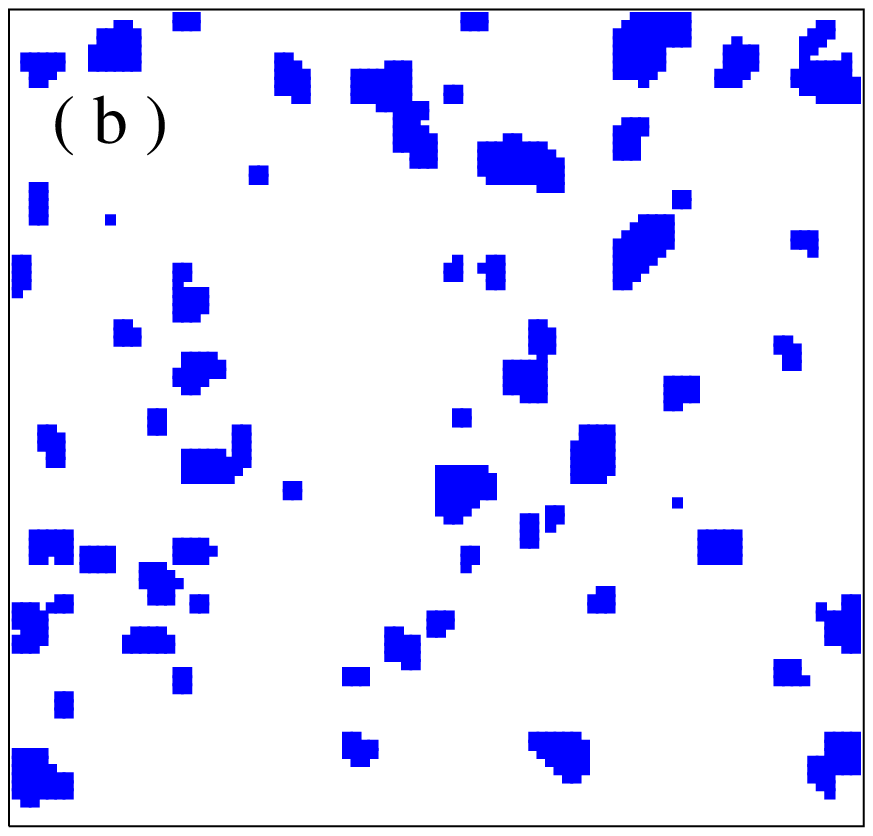}
\end{center} 
\caption{Crystal configurations under grinding (a) without
autocatalysis at a time $\nu_0 t=5600$, and (b) with autocatalysis at
 $\nu_0 t=2900$.
}
\label{fig3}
\end{figure}

\noindent
{\it With Grinding but Without Autocatalysis}

We then introduce the process of grinding. Grinding is performed at each
time interval of $\nu_0/f=1$. Since the crystal clusters are broken randomly,
they have small sizes as shown in Fig.3(a) at a time $\nu_0 t=5600$.
The cluster sizes do not increase, and seem to remain more or less 
a steady value already from the time  $\nu_0 t=100$ on.
Without autocatalysis, two enantiomers coexist and no chirality
selection takes place, as shown by the curve C1 in Fig.2(a). 
The dominant enantiomer changes ocasionally from R to S, or vice versa.
Even with a finite cee at the initial state, the value fluctuates 
but does not increase, as shown by the curve C2 in Fig. 2(a).\\

\noindent
{\it With Grinding and Autocatalysis}

If one add autocatalysis in addition to the grinding process
such that $\nu_n=10^n \nu_0$ for $n=1$ to 4, then the
selection of chirality takes place quickly, as shown 
by the curve D1 in Fig. 2(a).
During the initial incubation stage,
two enantiomers compete each other and the chirality $\phi$
changes sign randomly.
After the incubation time of about $\nu_0 t \approx 500$,
the homochiral state is achieved within a short time till $\nu_0 t=2800$;
only 2800 grinding processes are totally undertaken.
The cee order parameter $|\phi|$ increases exponentially, 
as shown in Fig. 2(b).
The asymptotic growth is fitted by 
$|\phi|=0.031 \exp(\nu_0 t/800).$
This exponential increase of cee is
in good agreement with the observation in the experiment\cite{noorduin+08b}.

When the initial state is a little chiral, 
as shown by the curve D2 in Fig. 2(a),
cee increases quickly to the final homochiral state.
By shifting the curve D2 parallel to the time axis, this curve 
overlaps well with the curve D1.

\begin{figure}[tb]
\begin{center} 
\includegraphics[width=0.45\linewidth]{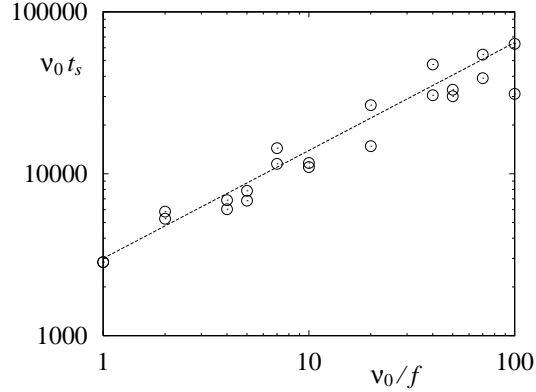}
\end{center} 
\caption{Saturation time $\nu_0 t_s$ versus inverse grinding frequency 
$\nu_0/f$ in both logarithmic represntation.
}
\label{fig4}
\end{figure}

\begin{figure}[tb]
\begin{center}
\includegraphics[width=0.45\linewidth]{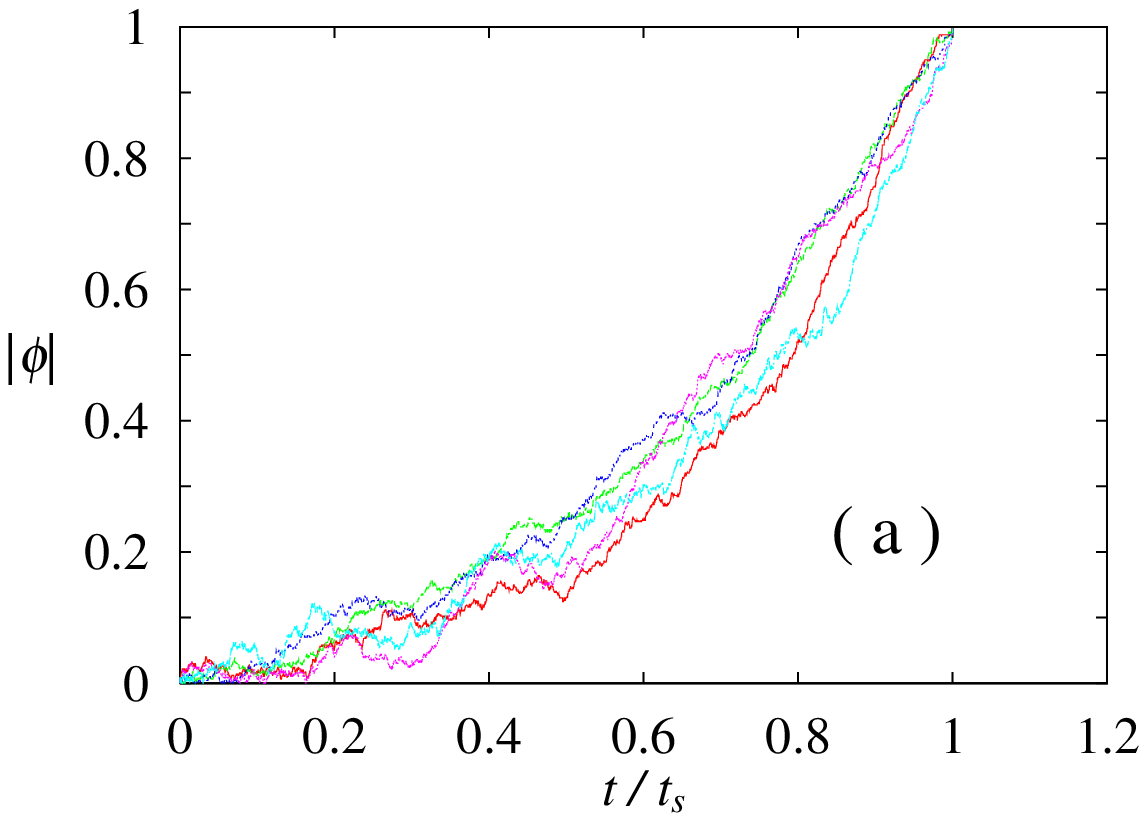}
\includegraphics[width=0.45\linewidth]{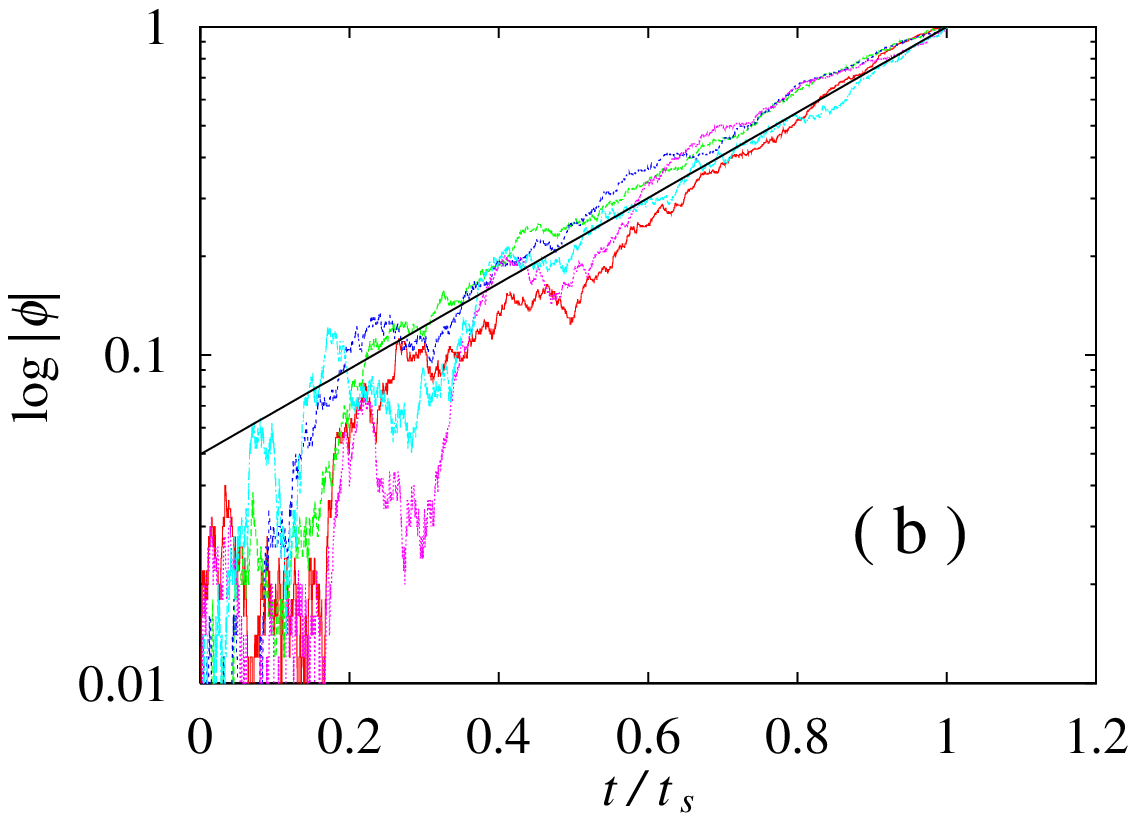}
\end{center}
\caption{Scaled time evolution of cee order parameter $|\phi|$
versus scaled time, $\tilde t=t/t_s$.
}
\label{fig5}
\end{figure}

We define the saturation time $t_s$ as the time 
necessary for the completion of homochirality: $|\phi(t_s)|=1$.
When the grinding frequency $f$ decreases, the saturation
time $t_s$ gets longer.
Even though there is a large fluctuation due to the stochastic character
of incubation time, saturation time $t_s$ can be fitted as 
$\nu_0 t_s \propto (\nu_0 /f)^{2/3}$, as shown in Fig. 4.
Because of our artificial grinding process, it is so far difficult to
explain this scaling behavior theoretically.
In any case, as the grinding rate $f$ diminishes, the necessary time for
the chiral symmetry breaking diverges. 
In fact, only with autocatalysis but without grinding, crystal
clusters grow in size as in the normal Ostwald ripening, but
cee remains very small, as shown by a pink curve B1 in Fig. 2(a), 
until the time 
$\nu_0 t= 7 \times 10^4$.
Grinding is essential for the
homochirality in crystal growth.

The temporal development of $|\phi|$ is found to be scaled
in terms of scaled time $t/t_s$.
In the scaling plot of Fig. 5(a), various simulations
 at different frequency $f/\nu_0$ between 1 and 0.01, are found to
 collapse well on a single line.
The semi-logarithmic plot in Fig. 5(b) indicates that 
$|\phi|$ increases exponentially as $|\phi| \sim e^{3(t/t_s-1)} $.\\

\noindent
{\it Conclusion}

We have extended the standard lattice gas model of crystal growth to
include the processes of molecular chirality conversion and grinding.
Simulation studies show that the chirality selection only
by Ostwald ripening without grinding is extremely slow, if possible. 

Grinding is simulated by cutting and shafling the system randomly,
but grinding alone cannot achieve chirality selection either.
For the accomplishment of homochirality, we need "autocatalytic effect" 
on crystallite surface,  in particular on kink sites, where the
 conversion rate of a molecular chirality is enhanced, as well
 as grinding. In this way, the exponential approach to homochiral state
 is realized.
  
Our model does not directly distribute monomers at the grinding process,
but breaking of crystal clusters creats many small fragments,
which quickly dissolves into monomers.
Monomers thus spread land on rough surface of relatively large
 crystal clusters, 
and thanks to the enhanced rate of chirality 
change on the crystal surface and kink sites, 
monomer's chirality conversion in an autocatalytic way. 
Ultimately, homochiral state
is realized very rapidly through this enhanced monomer recycling process.

 One has to note that this kind of  homochiral state 
 is not the equilibrium state.
 The latter is the one with only a single large crystal with a
 minimum surface free energy, and Ostwald ripening descibes
 the relaxation process to it:
Homochiral state with many small crystallites has 
 surely an excessive surface free energy.
Furthermore, a homochiral state which has configurations
similar to that shown in Fig.4b
has a higher free energy than the racemic state obtained
by changing the chirality of each crystallite randomly; 
The former state lacks the mixing entropy.
Thus, the homochiral state can only be selected when the system is
far away from equilibrium, for example driven by grinding in the
present case \cite{asakura+08}. 

For the growth of chiral crystals from achiral substrate, as NaClO$_3$, 
one has to include additional constituent, an achiral monomer A,
into the system. Instead of the direct chirality conversion
between R and S enantiomers in the solution,  chirality should appear 
only when molecules form clusters. At the incorporation stage
of an achiral molecule to the cluster, its chirality is determined.
When the conversion rate depends on the surrounding,
an effective autocatalytic process will be provoked.
This is the senario we study in future in relation to the Viedma's experiment
\cite{viedma04}. Extension of the model to three-dimensional case
is also an interesting future subject.

\acknowledgement
Y.S. acknowledges discussions with C. Viedma, W. Noorduin, M. Lahav,
M. McBride , A. Commeyras, A. Kagan, K. Soai, K. Asakura and
all the participants of the NORDITA workshop on "Origin of Chirality" 
in 2008, organized by A. Brandenburg and R. Plasson.
Y.S. acknowledges support by a Grant-in-Aid for Scientic Research 
(No. 19540410) from the Japan Society for the
Promotion of Science.


\end{document}